\documentclass[aps,prl,twocolumn,showpacs,superscriptaddress]{revtex4-1}
\usepackage{amsmath,amssymb}
\usepackage[pdftex]{graphicx}
\usepackage{booktabs}
\usepackage{color}
\newcommand \beq{\begin{eqnarray}}
\newcommand \eeq{\end{eqnarray}}
\newcommand \lk{\left(}
\newcommand \rk{\right)}
\begin{document}

\title{Anomalous Transport in a
  Superfluid Fluctuation Regime}
\author{Shun Uchino}
\affiliation{RIKEN Center for Emergent Matter Science, Wako,
  Saitama 351-0198, Japan}
\author{Masahito Ueda}
\affiliation{Department of Physics, University of Tokyo,
  7-3-1 Hongo, Bunkyo-ku, Tokyo 113-0033, Japan}
\affiliation{RIKEN Center for Emergent Matter Science, Wako,
  Saitama 351-0198, Japan}
\begin{abstract}
  Motivated by a recent experiment in ultracold atoms [
  S. Krinner et al., Proc. Natl. Acad. Sci. U.S.A \textbf{113}, 8144 (2016)], we
  analyze transport of attractively interacting fermions
  through a one-dimensional wire
  near the superfluid transition.
  We show that in a ballistic regime where the
  conductance is quantized 
  in the absence of interaction,
  the conductance is renormalized by
  superfluid fluctuations in reservoirs.
  In particular, the particle conductance
  is strongly enhanced and the plateau is blurred by
  emergent bosonic pair transport.
  For spin transport, in addition to the contact resistance
  the wire itself is resistive,
  leading to a suppression of the measured spin conductance.
  Our results are qualitatively consistent with the experimental observations.
\end{abstract}
\maketitle

Transport measurements often play crucial roles in revealing
the fundamental nature of matter.
In condensed matter physics,
superconductivity, the Kondo effect, and the quantum Hall effect
were all discovered with transport measurements.
A two-terminal setup realized in
ultracold atoms has opened up yet another avenue to explore
strongly correlated systems through
transport~\cite{Brantut2012aa,Stadler:2012kx,Krinner2013,
  Brantut:2013aa,Krinner:2015aa,
  Husmann2015aa,Krinner2016}.
In ultracold atoms,  quantum transport typically occurs in the clean limit. 
The bulk conductivity cannot
distinguish between different quantum states,
since the $f$-sum rule and the momentum conservation dictate
that the conductivity involves the delta-function 
singularity at zero frequency 
whose weight does not depend on detailed states of matter~\cite{Enss2012}.
However, transport through a constriction such as a quantum point contact
allows one to distinguish between different states 
due to the breakdown of the
momentum conservation at the constriction.
Indeed, such a setup has unveiled
different transport properties
for non-interacting~\cite{Krinner:2015aa} and superfluid
fermions~\cite{Husmann2015aa}.

Recently, particle conductance and spin conductance
have been measured with a quantum point contact
in ultracold fermions~\cite{Krinner2016}.
There, with increasing attractive interaction,
both of them deviate significantly from 
quantized values~\cite{Imry2002} just
above the superfluid critical temperature.
More specifically,
compared with the non-interacting
limit, the particle conductance 
is enhanced, 
whereas the spin conductance is suppressed.
These remarkable features stand in sharp contrast with the conventional wisdom
that a conductance is not renormalized by an
interaction in a one-dimensional
wire~\cite{Tarucha1995,Maslov1995,Ponomarenko1995,Safi1995,
  Kawabata1996,Shimizu1996}.
Meanwhile, different from condensed matter situations,
fluctuation and interaction effects in reservoirs may be
significant in cold atom experiments.

On another front, to cope with effects of an interaction at reservoirs
in ballistic transport presents a theoretical challenge, since
Landauer's approach does not operate with an interaction,
and phenomenological
approaches~\cite{Maslov1995,Ponomarenko1995,Safi1995,Kawabata1996,Shimizu1996}
used to
explain the interaction effect in a one-dimensional wire
cannot directly answer the question.
When the conductance of the wire is small, 
a tunneling Hamiltonian approach is widely used
to investigate the effect of  interactions in reservoirs
on transport~\cite{Mahan2013,Larkin2005,Varlamov1983,Levchenko2010}.
However, to discuss the ballistic limit realized in
Ref.~\cite{Krinner2016}, we must go beyond the linear response theory which has been 
widely used in
tunneling experiments with correlated
materials~\cite{Fischer2007}.

In this Letter, motivated by the ETH experiment~\cite{Krinner2016}
and the theoretical challenge mentioned above,
we examine the effects of superfluid fluctuations in reservoirs
on transport through a one-dimensional wire.
To deal with ballistic transport,
we apply the nonlinear response
theory~\cite{Wehrum1974,Genenko1986,Butcher1991,Rammer2007,Kamenev:2011aa}
to demonstrate that the breakdown of the quantization of conductance occurs by
superfluid fluctuations.
We show that transport of preformed pairs induced by superfluid fluctuations
is essential to account for the breakdown.
We also point out that in addition to the contact resistance,
the resistance in the one-dimensional wire plays an important role in
spin transport.

\textit{The Model.}---
We consider a system where
two macroscopic reservoirs with superfluid fluctuations
are connected by a quantum point contact (a one-dimensional channel).
In the ETH experiment, the constriction 
has potential variations that take place over length scales larger than $1/k_F$
with the Fermi momentum $k_F$. This implies that
when transport near the Fermi energy is concerned,
the adiabatic approximation is justified in which
the detailed shape in the constriction is irrelevant~\cite{Imry2002,suppl}.
Thus, for the single channel case,
we can start with the following Hamiltonian ($\hbar=k_B=1$):
\beq
&&H=\sum_{j=L, R}[\sum_{\mathbf{p}}\sum_{\sigma=\uparrow,\downarrow}
  \xi_{j,\mathbf{p},\sigma}
c^{\dagger}_{j,\mathbf{p},\sigma}
c_{j,\mathbf{p},\sigma}+V_j]+H_T, \\ 
&&V_j=
-g\sum_{\mathbf{p},\mathbf{p}',\mathbf{q}}
c^{\dagger}_{j,\mathbf{p+q},\uparrow}c^{\dagger}_{j,-\mathbf{p},\downarrow}
c_{j,-\mathbf{p}',\downarrow}c_{j,\mathbf{p'+q},\uparrow},\\
&&H_T=\int d\mathbf{x}d\mathbf{y}\sum_{\sigma}t(\mathbf{x},\mathbf{y})
\psi^{\dagger}_{L,\sigma}(\mathbf{x})\psi_{R,\sigma}(\mathbf{y})
+\mathrm{h.c.},
\eeq
where $c_{j,\mathbf{p},\sigma}$ $(c^{\dagger}_{j,\mathbf{p},\sigma})$  is
the fermionic
annihilation (creation) operator with momentum
$\mathbf{p}$ and spin $\sigma$ in reservoir $j$,
and
the energy
$\xi_{j,\mathbf{p},\sigma}=\frac{\mathbf{p}^2}{2m}-\mu_{j,\sigma}$ is
measured from the chemical potential $\mu_{j,\sigma}$.
In addition, $\psi_{j,\sigma}$ is the operator in the real space
with its argument
$\mathbf{x}$ ($\mathbf{y}$) representing a
position in the left (right) reservoir.
The first and second terms on the right-hand side of Eq.~(1) are
the single-particle Hamiltonian
and the interaction with an attractive coupling $-g$ ($g>0$),
respectively, and describe the system with a
broad-Feshbach resonance  used in
the ETH experiment~\cite{Krinner2016}.
Below, we focus on the BCS regime 
($1/(k_Fa)<0$ with the $s$-wave scattering length
$a$)~\cite{Haussmann2007,Bloch2008,levinsen2015,suppl}
above the superfluid transition temperature $T_c$.
To discuss the case of a single conducting channel, we set
$t(\mathbf{x},\mathbf{y})=t\delta(\mathbf{x}-\mathbf{x}_0)
\delta(\mathbf{y}-\mathbf{y}_0)$~\cite{suppl,Cuevas1996,Berthod2011},
where near the Fermi energy
the tunneling amplitude $t$ can be chosen to be a real constant without
loss of generality, and
$\mathbf{x}_0$ ($\mathbf{y}_0$) is the entrance (exit) point in the quantum
point contact.
In fact, this  tunneling Hamiltonian can precisely reproduce
the known \textit{universal}
conduction properties in the quantum point contact including 
ballistic transport with superfluid
reservoirs~\cite{Cuevas1996,Husmann2015aa}.

In terms of Eq.~(3), 
the mass and spin current operators are given by
$I_{\text{mass}}=-\sum_{\sigma}\dot{N}_{L,\sigma}=-\sum_{\sigma}i[H_T,N_{L,\sigma}]$,
and 
$I_{\text{spin}}=-i[H_T,N_{L,\uparrow}]+i[H_T,N_{L,\downarrow}]$,
where $N_{L,\sigma}=\int d\mathbf{x}\psi^{\dagger}_{L,\sigma}(\mathbf{x})
\psi_{L,\sigma}(\mathbf{x})$ is the number operator with spin $\sigma$ in the reservoir $L$.
We note that the number operator in each
reservoir commutes with the Hamiltonian except
for $H_T$.
\begin{figure}[t]
 \begin{center}
  \includegraphics[width=0.75\linewidth]{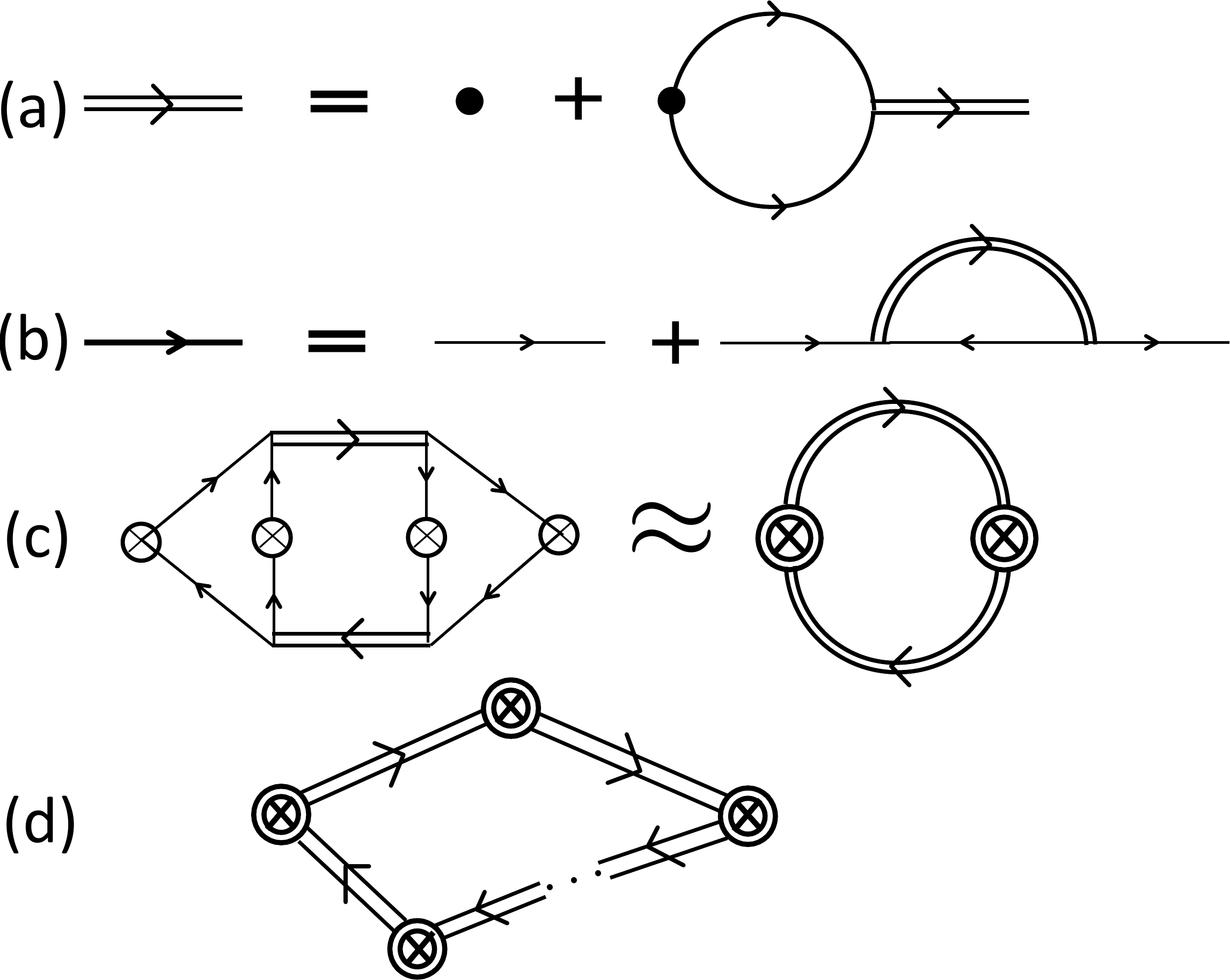}
  \caption{(a) Dyson's equation for a pair-fluctuation propagator
    $L(\mathbf{q},\omega)$.
    The dot and the solid line represent the interatomic coupling ($-g$) and
    the single-particle Green's function, respectively.
    (b) Diagrammatic representation of 
    the single-particle Green's function with the first-order correction
    of pair fluctuations represented by the double line.
  (c) Lowest-order diagram on the fluctuation pair tunneling.
  Each circle with a cross mark represents the tunneling amplitude $t$.
  This process can be replaced by the direct pair
  exchange diagram shown on the right
  with a renormalized tunneling amplitude $\tilde{t}$
  (double circle with a cross mark).
  (d) Higher-order diagram of the fluctuation pair tunneling.}
  \label{fig1}
 \end{center}
\end{figure}
In the presence of a chemical-potential difference between the reservoirs,
$V\equiv \mu_{L,\uparrow}-\mu_{R,\uparrow}=
\mu_{L,\downarrow}-\mu_{R,\downarrow}\neq0$
($V\equiv \mu_{L,\uparrow}-\mu_{R,\uparrow}=-\mu_{L,\downarrow}+\mu_{R,\downarrow}$),
the mass (spin) current is induced.
Then, the averages of the mass and spin currents at time $\tau$ are given by
\beq
I_{\text{mass/spin}}(\mathbf{x}_0,\mathbf{y}_0,\tau)=2\text{Im}\{e^{-iV\tau}
\langle A_{\uparrow}(\mathbf{x}_0,\mathbf{y}_0,\tau)\rangle_{H}\nonumber\\
\pm e^{\pm iV\tau}
\langle A_{\downarrow}(\mathbf{x}_0,\mathbf{y}_0,\tau)\rangle_{H}\},
\label{eq:current}
\eeq
where $A_{\sigma}(\mathbf{x}_0,\mathbf{y}_0,\tau)=t
\psi^{\dagger}_{R,\sigma}(\mathbf{y}_0,\tau)\psi_{L,\sigma}(\mathbf{x}_0,\tau)$,
and $\langle\cdots\rangle_{H}$ means the thermal average for
the Hamiltonian~(1).

In the presence of superfluid fluctuations, we should consider
contributions arising from
fermionic quasiparticles and fluctuation pairs~\cite{Larkin2005}.
Below, such fluctuations are considered
up to the Gaussian level in each propagator, which 
is reasonable in a regime $10^{-3}\lesssim(T-T_c)/T_c\lesssim10^{-1}$
for the case
of
three-dimensional reservoirs~\cite{Larkin2005} relevant to the ETH experiment.

\textit{Fermionic quasiparticle current.}---We
now examine a steady current induced by fermionic quasiparticles.
By the assumption of the steady state, we can put $\tau=0$ without loss of
generality. Then, the mass and spin currents can be expressed as
$I_{\text{mass/spin}}=\frac{t}{2\pi}\int d\omega
\text{Re}[G^{K}_{\uparrow}(\mathbf{x}_0,\mathbf{y}_0,\omega)
  \pm G^{K}_{\downarrow}(\mathbf{x}_0,\mathbf{y}_0,\omega)],$
where $G^{K}_{\sigma}(\mathbf{x}_0,\mathbf{y}_0,\omega)=
-i\int d\tau e^{i\omega \tau}\langle[\psi_{L,\sigma}(\mathbf{x}_0,\tau),
  \psi^{\dagger}_{R,\sigma}(\mathbf{y}_0,0)]\rangle_{H}$
is the Keldysh Green's function~\cite{Kamenev:2011aa}, and
we use $\int d\omega \text{Re}G^{R}_{\sigma}(\mathbf{x}_0,\mathbf{y}_0,\omega)=0$
for the retarded Green's function
$G^R_{\sigma}$~\cite{Rammer2007,Kamenev:2011aa,Berthod2011}.

As shown in Fig.~\ref{fig1}~(b),
the single-particle Green's function  is renormalized by
the fluctuation pair propagator (Fig.~\ref{fig1}~(a)) up to the Gaussian level.
As pointed out in Ref.~\cite{Cuevas1996}, 
the effect of $t$ must be incorporated to all orders in the ballistic limit.
By using an analysis similar to the case of noninteracting
fermions~\cite{Genenko1986,Caroli1971,Cuevas1996,Berthod2011}
the fermionic quasiparticle contribution to the conductance per spin is obtained as
\cite{suppl}
\beq
G_{q}\approx\frac{1}{h}\int_{-\infty}^{\infty} d\omega \frac{4\pi^2 t^2\rho^2(\omega)}{
  |1+\pi^2t^2\rho^2(\omega)|^2
  }\lk-\frac{\partial n_{F}(\omega)}{\partial \omega}\rk,
\label{eq:qcurrent}
\eeq
where $n_{F}(\omega)=(e^{\omega/T}+1)^{-1}$  
is the Fermi distribution function at temperature $T$,
$g^R(\omega)=\sum_{\mathbf{p}}g^{R}(\mathbf{p},\omega)$
with the retarded Green's function $g^{R}(\mathbf{p},\omega)$
in the reservoirs, and
$\rho(\omega)=-\text{Im}[g^{R}(\omega)]/\pi$
is the density of state (DOS).
We note that the conductance depends neither on $\mathbf{x}_0$ nor
$\mathbf{y}_0$~\cite{suppl}.
To obtain the above result, an expansion up to
linear order in $V$  is considered,
since $V/\mu_{L(R)}\lesssim0.1$ and no significant deviation from the linear order 
is found in the ETH experiment.
We note that
the same expression holds for the mass and spin currents.
In the case of small transmittance where
$|1+\pi^2t^2\rho^2(\omega)|^2\approx1$ in the denominator,
 Eq.~\eqref{eq:qcurrent} essentially reduces
to the Ambegaokar-Baratoff formula~\cite{Ambegaokar1963,Larkin2005}.
On the other hand, in the absence of the fluctuations,
the conductance is reduced to
$
G_{q}= \frac{T_0}{h},
$
and is equivalent to Landauer's formula
with transmittance $T_0=4\pi^2t^2\rho_0^2(0)/(1+\pi^2t^2\rho_0^2(0))^2$,
where
$\rho_0$ is the DOS for noninteracting
fermions and we use the fact that the change of $\rho_0$ around the Fermi level
is much smaller than that of $\frac{\partial n_F}{\partial \omega}$
~\cite{Genenko1986,Caroli1971,Cuevas1996}.
In this limit, the quantized conductance, $1/h$, is obtained for
$t=1/(\pi\rho_0(0))$.

The superfluid fluctuations renormalize
the conductance of fermionic quasiparticles and
generate that of preformed pairs. The fermionic quasiparticle conductance,
in general, tends to be suppressed due to pseudogap effect~\cite{Janko1997,Tsuchiya2009};
however, in the case of three-dimensional reservoirs, such suppression is found
to be negligible in the experimentally relevant regime $(T-T_c)/T_c\sim10^{-1}$~\cite{suppl}.

\textit{Fluctuation pair current.}---We now 
consider a  current carried by the fluctuating (preformed) pairs
that makes a dominant contribution to the conductivity
in dirty superconductors~\cite{Larkin2005}.
As shown on the left of
Fig.~\ref{fig1}~(c), the lowest-order diagram  already
contains a factor
$t^4$. In usual tunneling experiments
where
$\pi t\rho_0(0)\ll1$~\cite{Mahan2013,Larkin2005,Varlamov1983,
  Levchenko2010,Fischer2007},
this contribution is negligible compared with the fermionic
quasiparticle current,
and has not been considered for realistic situations.
However, in the ballistic regime in which $\pi t\rho_0(0)\approx1$,
one needs to consider it seriously.
To evaluate the fourth-order diagram,
we calculate the
third-order response function, which is related to
an imaginary-time correlation  function through analytic
continuation~\cite{Mahan2013,Wehrum1974}.
Up to linear order in $V$, the fluctuation pair current
at the fourth order in $t$ behaves as
$I^{(4)}_p\sim t^4(2V)/(T-T_c)$.
Here, the factor $2V$
originates from the pair exchange between the reservoirs, and
the factor $1/(T-T_c)$ reflects the superfluid fluctuations.
We note that  as in the case of spin conductivity~\cite{Enss2012},
the fluctuation pair contribution in the spin current
vanishes.
This reflects the fact that the pair exchange is not
caused by a spin bias.
Thus, the enhancement of the current by fluctuation pairs
only occurs for mass transport.

We also note that the left-hand side of
Fig.~\ref{fig1}~(c)
can be replaced by the right-hand side of Fig.~\ref{fig1}~(c) up to linear order in $V$.
Namely, the fluctuation pair contribution can be expressed
in terms of an effective hopping amplitude 
$\tilde{t}=(\pi t\rho_0(0))^2/(2T)$
\footnote{Notice that the dimensional of $\tilde{t}$ is different from that of $t$, since
the dimension of $L^R$ is different from that of $g^R$.} and
the retarded pair-fluctuation  propagator
whose expression in the vicinity of
$T_c$ is given by \cite{Larkin2005}
\beq
L^{R}(\mathbf{q},\omega)=\frac{8T}{\pi \rho_0(0)}
\frac{1}{i\omega-(\tau_{\text{GL}}^{-1}+\frac{8T\xi^2}{\pi}q^2)},
\eeq
 where
$\tau_{\text{GL}}^{-1}=8(T-T_c)/\pi$ and $\xi^2=7\zeta(3)v_F^2/(16d\pi^2T^2)$
with the Fermi velocity
$v_F$ and the dimension of the system $d$.
Thus, this contribution can be calculated
as tunneling of the preformed pairs for a given
effective hopping amplitude and bias $2V$, and therefore
 the multiple tunneling processes of the preformed pairs
represented by power series in $\tilde{t}$
can be systematically evaluated as depicted in Fig.~\ref{fig1}~(d).
By using the nonlinear response
theory, we obtain the conductance contributed from
the fluctuation pair  per spin up to linear order in $V$ as~\cite{suppl}
\beq
G_p\approx\frac{1}{h}\int_{-\infty}^{\infty}
\frac{d\omega}{\sinh^2(\frac{\omega}{2T})}
     \frac{\frac{2\tilde{t}^2}{ T}(\sum_{q}
       \text{Im}[L^{R}(q,\omega)])^2}{
       \{1-\tilde{t}^2(\sum_{q}
       \text{Re}[L^{R}(q,\omega)])^2\}^2}.\nonumber\\
\eeq
We note that the above formula indeed reflects bosonic transport,
since the term $1/\sinh^2(\frac{\omega}{2T})$ in the integrand
is the derivative of the Bose distribution
function with respect to $\omega$ (note that such a term is absent in Eq.~(5)).
Thus, fluctuation pairs make a positive contribution
to the mass conductance.
Such a contribution is already visible in the regime
$(T-T_c)/T_c\sim10^{-1}$~\cite{suppl}.

\begin{figure}[t]
 \begin{center}
   \includegraphics[width=0.75\linewidth]{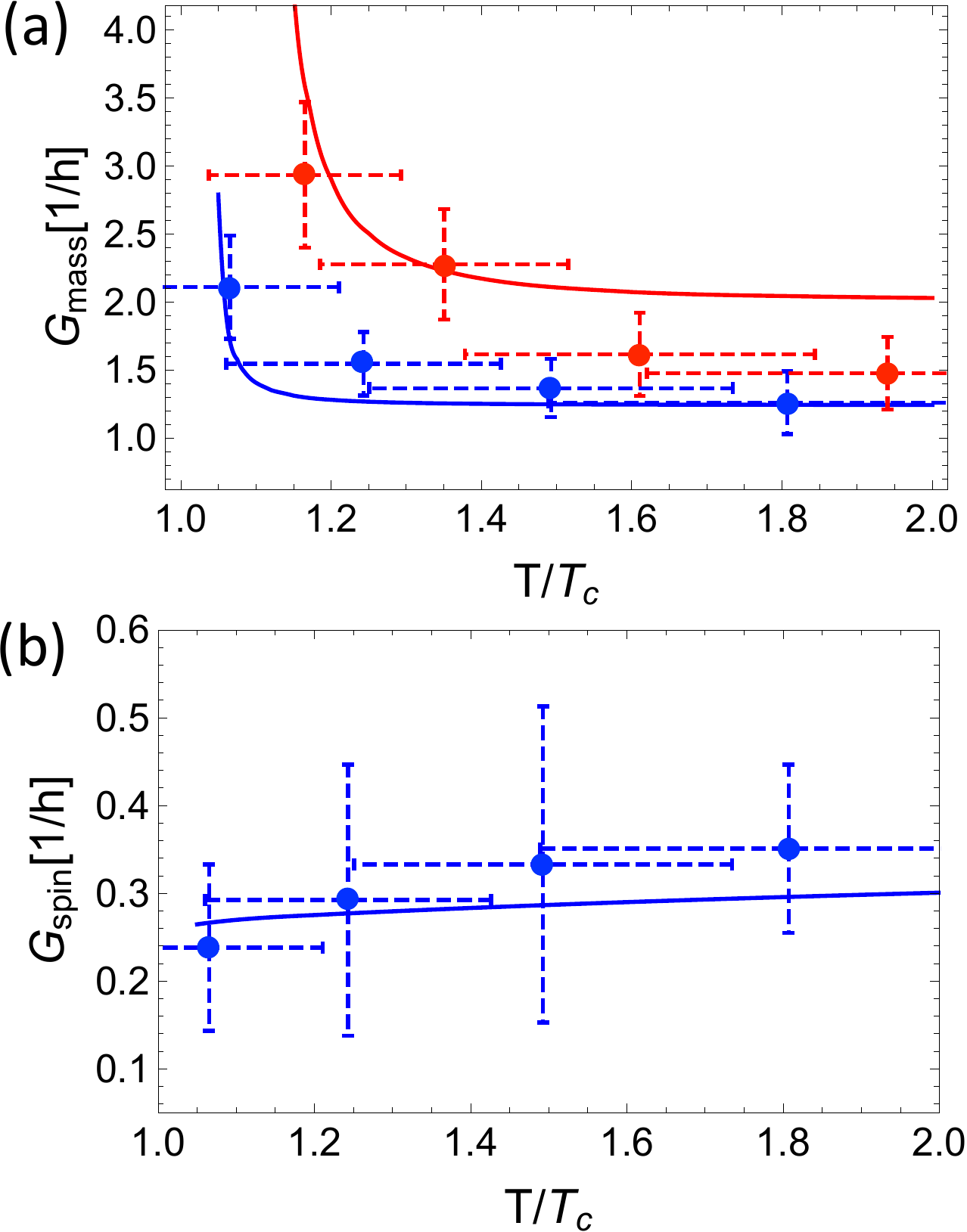}
   \caption{Particle conductance $G_{\text{mass}}$ (a) and
     the spin conductance $G_{\text{spin}}$ (b)
     as a function of $T/T_c$ in the single-mode regime.
     Circles with error bars and solid curves represent the ETH experimental
     data~\cite{Krinner2016}
     and our
     theoretical calculations, respectively.
     The  blue and red colors show the experimental results 
      with $T/T_F=0.075$ and
     $0.1$, respectively.
     In (b) the 
     wire resistance is estimated so as to be compatible with the experiment
     by  the relation $R_s/h\sim  e^{\Delta_s/T}$.}
  \label{fig2}
 \end{center}
\end{figure}

\textit{Conductances in the single-channel regime.}---We now
compare our theory with the ETH experiment in the ballistic single-channel regime.
To this end, one may also consider an interaction effect inside the wire.

For mass transport,
the mass current operator commutes with the bulk Hamiltonian containing
an interaction
in the wire~\cite{Enss2012,suppl}~(except, of course, for the tunneling term),
the wire resistance is expected to be negligible, and
the conductance calculation obtained above is directly applicable
to the ETH experiment.
An essential input parameter in the theory is
the ratio $T/T_F$ with the Fermi temperature $T_F$,
which is extracted from
the experiment~\cite{Krinner2016,suppl}.
By assigning the ballistic limit $\pi t\rho_0(0)=1$,
we compare the theory with the experiment
and find excellent agreement
as shown in
Fig.~\ref{fig2}~(a).
A crucial point here is that  the conductance is enhanced due to
the bosonic fluctuation-pair contribution.
Since our theory is based on an expansion from $T_c$,
some deviation is expected at $T/T_c\gtrsim2$.

For spin transport, the spin current operator and the Hamiltonian do
not commute even in the absence of the tunneling term,
giving rise to the wire resistance~\cite{Sommer2011,Sommer2011b,Duine2010,
  Bruun2011,Bruun2011b,Enss2012}.
In the presence of an attractive interaction,
a spin gap, $\Delta_s$ shows up.
A typical estimation suggests $10\text{nK}\lesssim\Delta_s\lesssim500\text{nK}$,
where the lower bound is estimated with
the Yang-Gaudin model at the density $n\sim10^{6}{/m}$
and the upper bound is determined from the binding energy of the confinement-induced
resonance $\sim0.6\hbar\omega_{\perp}$, where
$\omega_{\perp}$ is the transverse confinement
frequency~\cite{Olshanii1998,Fuchs2004}.
In Fig.~\ref{fig2}~(b), we show 
the spin conductance  
$G_{\text{spin}}$
whose resistance is the sum of the contact resistance and wire resistance. 
The wire resistance is
estimated so as to be compatible with the experiment by
assuming 
$R_s/h\sim e^{\Delta_s/T}$~\cite{Giamarchi2004}. 
Our result shows that the wire resistance for spin transport is 
of the order of the contact resistance,
implying that even in the ballistic limit
a nonnegligible chemical potential drop occurs inside the wire
due to the interaction between $\uparrow$ and $\downarrow$ spin components.

We also comment on an effect of the spin gap near the contacts.
The spin gap  in the wire
originates from the strong nesting effect allowed in a one-dimension
system~\cite{Giamarchi2004}.
On the other hand, near the contact,
multiple channels that render the spin gap smeared out through the
dimensional crossover are present.
Thus, the effect of the spin gap near the contacts is expected to make negligible
contributions to the contact resistance.

\begin{figure}[t]
 \begin{center}
   \includegraphics[width=0.75\linewidth]{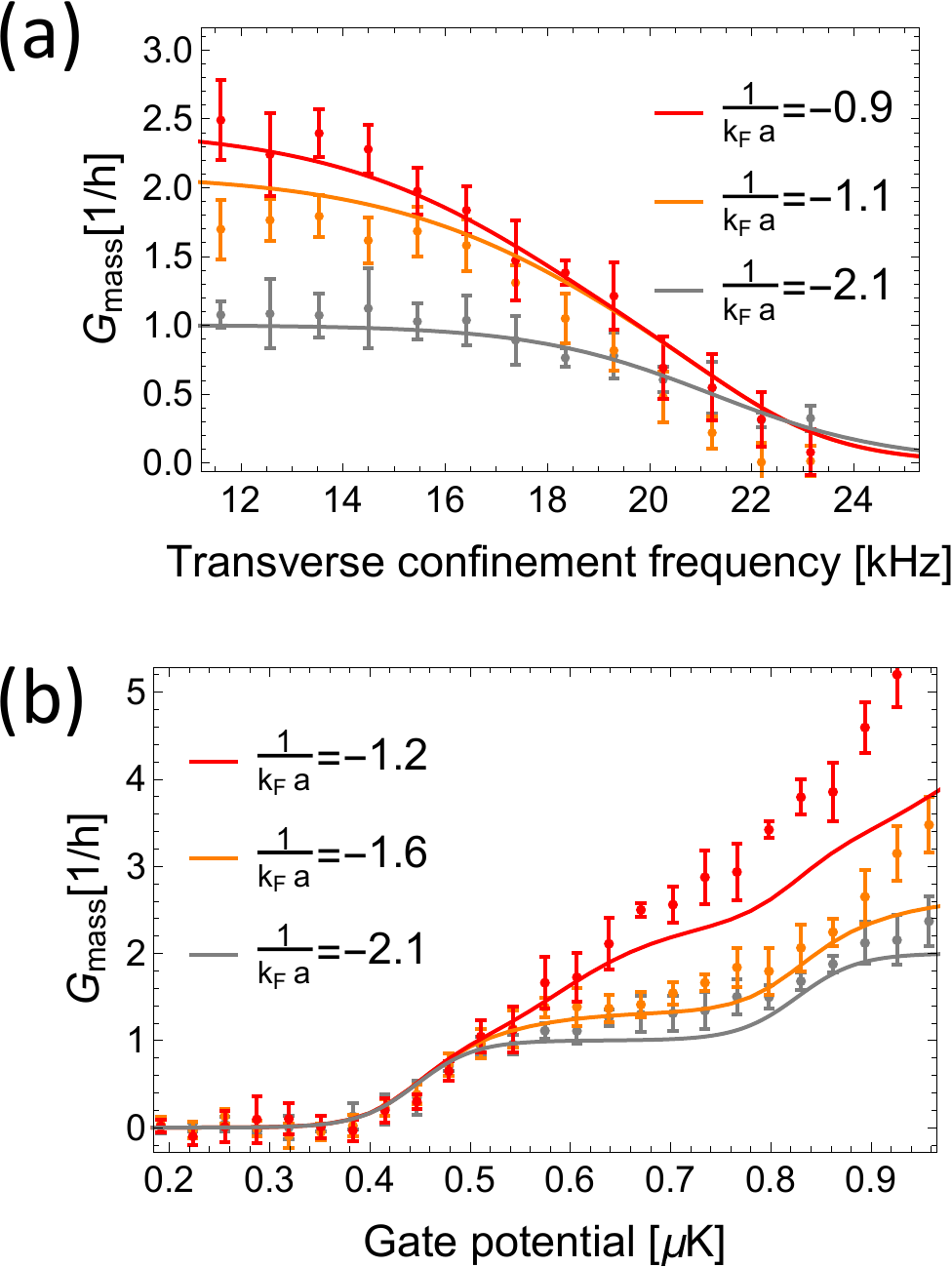}
   \caption{Comparison of the present theory with the ETH experiment
     in the particle conductance for different interaction strengths $1/(k_Fa)$.
     Conductances as a function of the transverse confinement frequency
     of the quantum
     point contact (a) and as a function of the gate potential (b).
     Solid curves are theoretical predictions, and
     circles with error bars represent experimental data whose color
     is identical to that of the corresponding theory curve.
     Grey curves are obtained with Landauer's formula
     for noninteracting fermions~\cite{suppl}.
     The parameter $T/T_F =0.1$ in Fig.~3~(a) and $T/T_F =0.075$ in Fig.~3~(b).
       }
  \label{fig3}
 \end{center}
\end{figure}

\textit{Effects of the gate potential, trapping, and interaction
  on particle conductance}---Now,
we discuss how the particle conductance is affected by 
the gate potential, trap potential, and interaction.
Since the gate and trap potentials
shift the energy levels of the conducting channels,
these effects can be incorporated
by introducing multiple tunneling amplitudes,
each of which depends on the gate and trap potentials~\cite{suppl,
  glazman1988,Buttiker1990,Krinner2016}.
The tunneling amplitudes as the input parameters
are determined so as to reproduce the weakest-interaction data
in the experiment based on Landauer's formula for noninteracting fermions,
since there, $(T-T_c)/T_c>1$ and the superfluid fluctuations
are expected to be minuscule~\cite{suppl}.
On the other hand, the interaction strength $1/(k_Fa)$
is directly related to
how close the system is to $T_c$, since
increasing $1/(k_Fa)$ towards the unitarity leads to an enhancement of $T_c$
\cite{Krinner2016,suppl}.

Figure~\ref{fig3} compares the results of our theory with the ETH experiments
for different interaction strengths
($T/T_F=0.1$ in Fig.~3~(a) and $T/T_F=0.075$ in Fig.~3~(b)~\cite{Krinner2016}).
For the weakest interaction $1/(k_Fa)=-2.1$,
the grey curves are obtained by
fittings of the experimental data by
assuming Landauer's formula.
For stronger interaction strengths, where $(T-T_c)/T_c<1$,
we use our theory by incorporating the
superfluid fluctuations to calculate the particle conductance
by assigning tunneling amplitudes estimated from the data at
$1/(k_Fa)=-2.1$.
As shown in Fig.~\ref{fig3},
the particle conductance is enhanced and
deviates from 
Landauer's formula, which is consistent with the experiment~\cite{Krinner2016}.
As in the case of the single-channel regime, enhancement is caused by 
the preformed pairs.
The discrepancy between the theory and
the experiment occurring at the larger gate potential
in Fig.~\ref{fig3}~(b)
may be due to an effect of the higher
transverse channels that are not treated in
the theory.

\textit{Summary.}---We have shown that superfluid fluctuations
cause two competing effects in two-terminal transport
through a quantum point contact;
suppression of the conductance of
fermionic quasiparticles and enhancement due to
bosonic preformed pairs.
The former is negligible in the ETH experiment,
since the depletion of the DOS near the Fermi level is 
negligible. The latter in the ballistic regime
is shown to be significant due to the absence of the Pauli exclusion
principle. Thus, the net conductance can exceed
the upper bound of the quantized conductance $1/h$ for noninteracting fermions 
through multiple tunneling processes that  
are captured with the nonlinear response
theory.
Such transport is ideally realized in an impurity-free system with
perfect transmittance such as cold atoms and high-mobility semiconductors
in which diffusive properties in a one-dimensional wire, which tends to suppress
the bosonic transport, can be ignored.
We have also shown that spin transport is affected by the wire resistance
originating from the spin gap whose determination
with no ambiguity requires the more precise
knowledge of the particle density in the wire.

\textit{Acknowledgement.}---We thanks Jean-Philippe Brantut, Tilman Esslinger,
Sebastian Krinner, Martin Lebrat, and
Naoto Tsuji for useful discussions.
This work is supported by
KAKENHI Grant Nos.~JP26287088 and
JP15H05855from the Japan Society for the Promotion of Science,
the Photon Frontier Network Program from MEXT of Japan,
and Grant for Basic Science Research Projects from The Sumitomo
Foundation.

\textit{Note added.}---Recently,
we became aware of works by
M. Kan{\'a}sz-Nagy et al.~\cite{Nagy2016} and
B. Liu et al.~\cite{Liu2016}, both of
which discuss the anomalous conductances
from different perspectives.

%

\end{document}